\documentclass[12pt]{article}

\usepackage{amsmath,amssymb,amsfonts,amsbsy}
\usepackage{graphicx}
\usepackage{wrapfig}

\begin{document}
\addcontentsline{toc}{subsection}{{Title of the article}\\
{\it B.B. Author-Speaker}}

\setcounter{section}{0}
\setcounter{subsection}{0}
\setcounter{equation}{0}
\setcounter{figure}{0}
\setcounter{footnote}{0}
\setcounter{table}{0}

\begin{center}
\textbf{STATISTICAL DESCRIPTION OF THE FLAVOR STRUCTURE OF THE NUCLEON SEA}\\
\vspace{3mm}
Jacques Soffer

\vspace{1mm}

\begin{small}
   \emph{Department of Physics, Temple University Philadelphia, Pennsylvania 19122-6082, USA}\\
   \emph{E-mail: jacques.soffer@gmail.com}\\
\end{small}
\vspace{3mm}
Claude Bourrely

\vspace{1mm}

\begin{small}
   \emph{ Aix-Marseille Universit\'e, D\'epartement de Physique, Facult\'e des Sciences de Luminy,\\
			F-13288 Marseille, Cedex 09, France}\\
   \emph{E-mail: claude.bourrely@univ-amu.fr}\\
\end{small}
\vspace{3mm}
Franco Buccella

\vspace{1mm}

\begin{small}
   \emph{ INFN, Sezione di Napoli, Via Cintia, Napoli, I-80126, Italy}\\
   \emph{E-mail: buccella@na.infn.it}
\end{small}	

\end{center}


\begin{abstract}
The theoretical foundations of the quantum statistical approach to parton distributions are reviewed together with the phenomenological motivations from a few specific features of Deep Inelastic Scattering data. The chiral properties of QCD lead to strong relations between quarks and antiquarks distributions and automatically account for the flavor and helicity symmetry breaking of the sea. We are able to describe both unpolarized and polarized structure functions in terms of a small number of parameters. The extension to include their transverse momentum dependence will be also briefly considered.
\end{abstract}

\vspace{6.2mm} 
\section{Basic review on the statistical approach}
Let us first recall some of the basic components for building up the parton
distribution functions (PDF) in the statistical approach, as oppose to the
standard polynomial type parametrizations, based on Regge theory at low $x$ and counting rules at large $x$. The fermion distributions are given by the sum of two terms \cite{bbs1}, the
first one, 
a quasi Fermi-Dirac function and the second one, a flavor and helicity
independent diffractive
contribution equal for light quarks. So we have, at the input energy scale
$Q_0^2$,
\begin{equation}
xq^h(x,Q^2_0)=
\frac{AX^h_{0q}x^b}{\exp [(x-X^h_{0q})/\bar{x}]+1}+
\frac{\tilde{A}x^{\tilde{b}}}{\exp(x/\bar{x})+1}~,
\label{eq1}
\end{equation}
\begin{equation}
x\bar{q}^h(x,Q^2_0)=
\frac{{\bar A}(X^{-h}_{0q})^{-1}x^{\bar b}}{\exp [(x+X^{-h}_{0q})/\bar{x}]+1}+
\frac{\tilde{A}x^{\tilde{b}}}{\exp(x/\bar{x})+1}~.
\label{eq2}
\end{equation}
It is important to remark that $x$ is indeed the natural variable, since all sum we will use are expressed in terms of $x$.
Notice the change of sign of the potentials
and helicity for the antiquarks.
The parameter $\bar{x}$ plays the role of a {\it universal temperature}
and $X^{\pm}_{0q}$ are the two {\it thermodynamical potentials} of the quark
$q$, with helicity $h=\pm$. We would like to stress that the diffractive
contribution 
occurs only in the unpolarized distributions $q(x)= q_{+}(x)+q_{-}(x)$ and it
is absent in the valence $q_{v}(x)= q(x) - \bar {q}(x)$ and in the helicity
distributions $\Delta q(x) = q_{+}(x)-q_{-}(x)$ (similarly for antiquarks).
The {\it nine} free parameters \footnote{$A$ and $\bar{A}$ are fixed by the following normalization conditions $u-\bar{u}=2$, $d-\bar{d}=1$.} to describe the light quark sector ($u$ and $d$), namely $X_{u}^{\pm}$, $X_{d}^{\pm}$, $b$, $\bar b$, $\tilde b$, $\tilde A$ and $\bar x$ 
in the above expressions, were
determined at the input scale from the comparison with a selected set of
very precise unpolarized and polarized Deep Inelastic Scattering (DIS) data
\cite{bbs1}. The additional factors $X_{q}^{\pm}$ and $(X_{q}^{\pm})^{-1}$ come from the transverse momentum dependence (TMD), as explained in Refs.~\cite{bbs6,bbs5} (See below). For the gluons we consider the black-body inspired expression
\begin{equation}
xG(x,Q^2_0)=
\frac{A_Gx^{b_G}}{\exp(x/\bar{x})-1}~,
\label{eq5}
\end{equation}
a quasi Bose-Einstein function, with $b_G$, the only free parameter, since $A_G$ is determined
by the momentum sum rule. We also assume a similar expression for the polarized gluon 
distribution $x\Delta G(x,Q^2_0)={\tilde A}_Gx^{{\tilde b}_G}/[\exp(x/\bar{x})-1]$. For the strange quark
distributions, the simple choice made in Ref.~\cite{bbs1}
was greatly improved in Ref.~\cite{bbs2}. Our procedure allows to construct simultaneously the unpolarized quark distributions and the helicity distributions. This is worth noting because it is a very unique situation. Following our first
paper in 2002, new tests against experimental (unpolarized and
polarized) data turned out to be very satisfactory, in particular in hadronic
collisions, as reported in Refs.~\cite{bbs3,bbs4}.

\section{Some selected results}

Let us first come 
back to the important question of the flavor asymmetry of the light
antiquarks. Our determination of $\bar u(x,Q^2)$ and
$\bar d(x,Q^2)$ is perfectly consistent with the violation of the Gottfried
sum rule, for which we found $I_G= 0.2493$ for $Q^2=4\mbox{GeV}^2$.
Nevertheless there remains an open problem with the $x$ distribution
of the ratio $\bar d/\bar u$ for $x \geq 0.2$.
According to the Pauli principle this ratio should be above 1 for any value of
$x$. However, the E866/NuSea Collaboration \cite{E866} has
released the final results corresponding to the analysis of their full
data set of Drell-Yan yields from an 800 GeV/c proton beam on hydrogen
and deuterium targets and they obtain the ratio, for $Q^2=54\mbox{GeV}^2$, 
$\bar d/\bar u$ shown in Fig. 1 (Left). 
Although the errors are rather large in the high $x$ region,
the statistical approach disagrees with the trend of the data.
Clearly by increasing the number of free parameters, it
is possible to build up a scenario which leads to the drop off of
this ratio for $x\geq 0.2$.
For example this was achieved in Ref. \cite{Sassot}, as shown 
by the dashed curve in Fig. 1 (Left). There is no such freedom in the statistical
approach, since quark and antiquark distributions are strongly related.
One way to clarify the situation is, to improve the statistical
accuracy on the Drell-Yan yields which seems now possible, since there are new opportunities
for extending the measurement of the $\bar {d}(x)/\bar {u}(x)$ ratio to larger $x$ up to $x=0.7$,
with the ongoing E906 experiment at the 120 GeV Main Injector at FNAL \cite{E906} and a proposed
experiment at the new 30-50 GeV proton accelerator at J-PARC \cite{J-PARC}.
\begin{figure}[ht]
\begin{center}
\includegraphics[width=6.9cm]{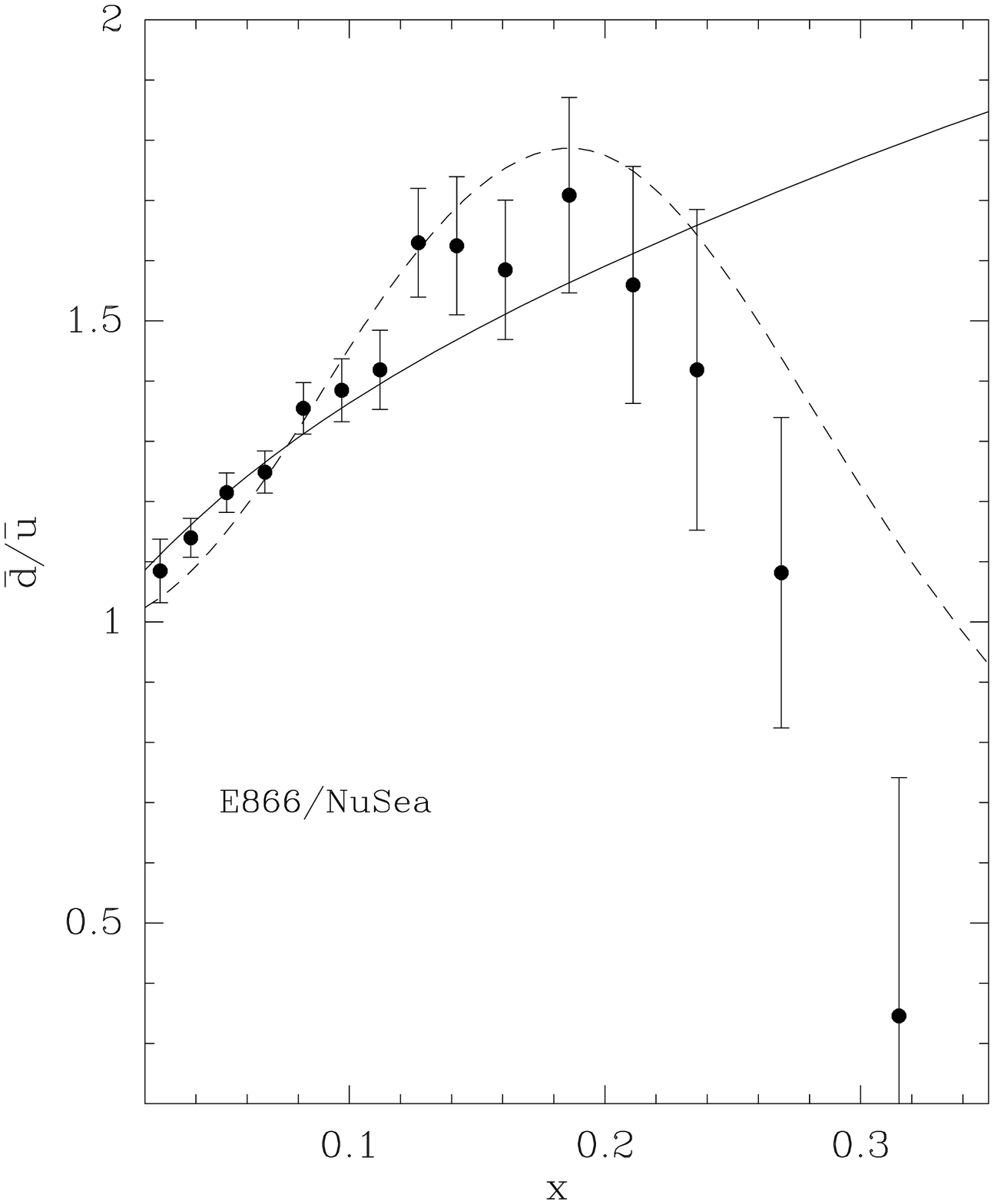}
\vspace*{10mm}
\hspace*{10mm}
\includegraphics[width=6.5cm]{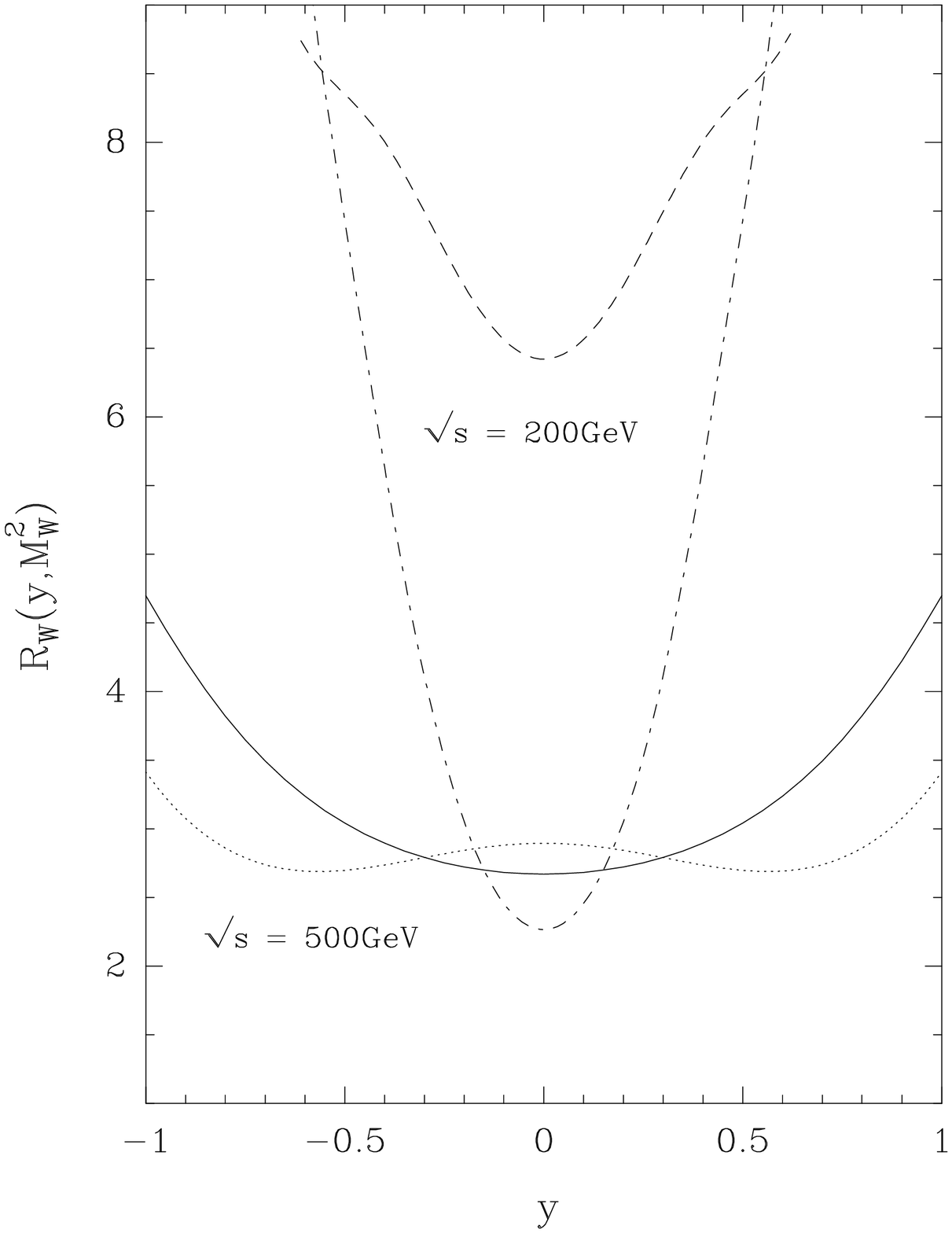}
\end{center}
\vspace*{-8mm}
\caption{{\it Left}: Comparison of the data on $(\bar d / \bar u) (x,Q^2)$  from E866/NuSea
at $Q^2=54\mbox{GeV}^2$
\cite{E866}, with the prediction of the statistical model (solid curve) 
and the set 1 of the parametrization proposed in Ref. \cite{Sassot}
(dashed curve). {\it Right}: Theoretical calculations for the ratio $R_W(y,M_W^2)$ versus the $W$ rapidity,
at two RHIC-BNL energies. 
Solid curve ($\sqrt s = 500\mbox{GeV}$) and dashed curve 
($\sqrt s = 200\mbox{GeV}$) are the statistical model predictions. 
Dotted curve ($\sqrt s = 500\mbox{GeV}$) and dashed-dotted curve 
($\sqrt s = 200\mbox{GeV}$) 
are the predictions obtained using the $\bar d(x) / \bar u(x)$ ratio
from Ref. \cite{Sassot}.
}
\label{fi:fig1}
\end{figure}

Another way is to call
for the measurement of a  different observable sensitive to
$\bar u(x)$ and  $\bar d(x)$.
One possibility is the ratio of the unpolarized cross sections for the
production of $W^+$ and $W^-$ in $pp$ collisions, which will directly probe
the behavior of the $\bar d(x) / \bar u(x)$ ratio.
Let us recall that if we denote 
$R_W(y)=(d\sigma^{W^+}/dy)/(d\sigma^{W^-}/dy)$, where $y$ is the
$W$ rapidity,  we have \cite{BSc} at the lowest order
\begin{equation}
R_W(y,M_W^2)= \frac{u(x_a,M_W^2) \bar d(x_b,M_W^2) + \bar d(x_a,M_W^2)
u(x_b,M_W^2)}{d(x_a,M_W^2) \bar u(x_b,M_W^2)
+ \bar u(x_a,M_W^2) d(x_b,M_W^2)}~,
\label{24}
\end{equation}
where $x_a=\sqrt{\tau}e^y$, $x_b=\sqrt{\tau}e^{-y}$ and $\tau=M_W^2/s$.
This ratio $R_W$, such that $R_W(y)=R_W(-y)$, is accessible 
with a good precision at RHIC-BNL \cite{BSSW} and at 
$\sqrt s = 500\mbox{GeV}$ for $y=0$, we have $x_a=x_b=0.16$.
So $R_W(0,M_W^2)$ probes the $\bar d(x) / \bar u(x)$ ratio at $x=0.16$. 
Much above this $x$ value, the accuracy of Ref. \cite{E866} becomes poor.
In Fig. 1 (Right) we compare the results for $R_W$ using two different
calculations.
In both cases we take the $u$ and $d$ quark distributions
obtained from the present analysis, but first we use the $\bar u$ and
$\bar d$ distributions of the statistical approach
(solid curve in Fig. 1 (Right)) and second the $\bar u$ and $\bar d$ from
Ref. \cite{Sassot} (dashed curve in Fig. 1 (Right)). For $y=\pm 1$, which corresponds to $x_a$ or $x_b$ near 0.43, one sees that
the predictions are very different.
Notice that the energy scale $M_W^2$ is much higher than in the E866/NuSea
data, so one has to take into account the $Q^2$ evolution. 
At $\sqrt s = 200\mbox{GeV}$ for $y=0$, we have $x_a=x_b=0.40$ and, 
although the 
$W^{\pm}$  yield is smaller at this energy, the effect on $R_W(0,M_W^2)$ is
strongly enhanced, as seen in Fig. 1 (Right). This is an excellent test, which needs to be revisited and should be done in the near future.
\begin{figure}[ht]
\begin{center}
\includegraphics[width=6.5cm]{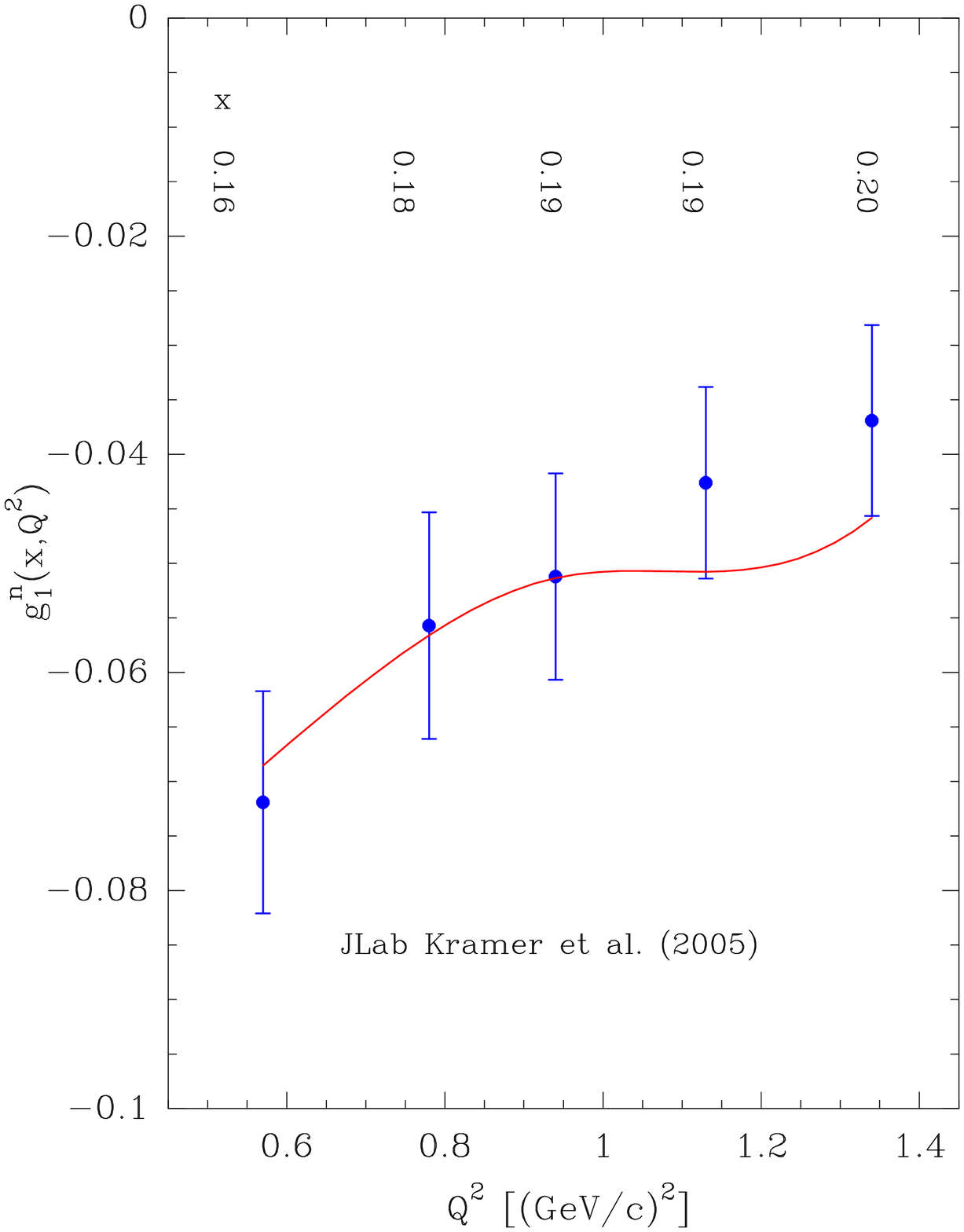}
\vspace*{10mm}
\hspace*{10mm}
\includegraphics[width=6.5cm]{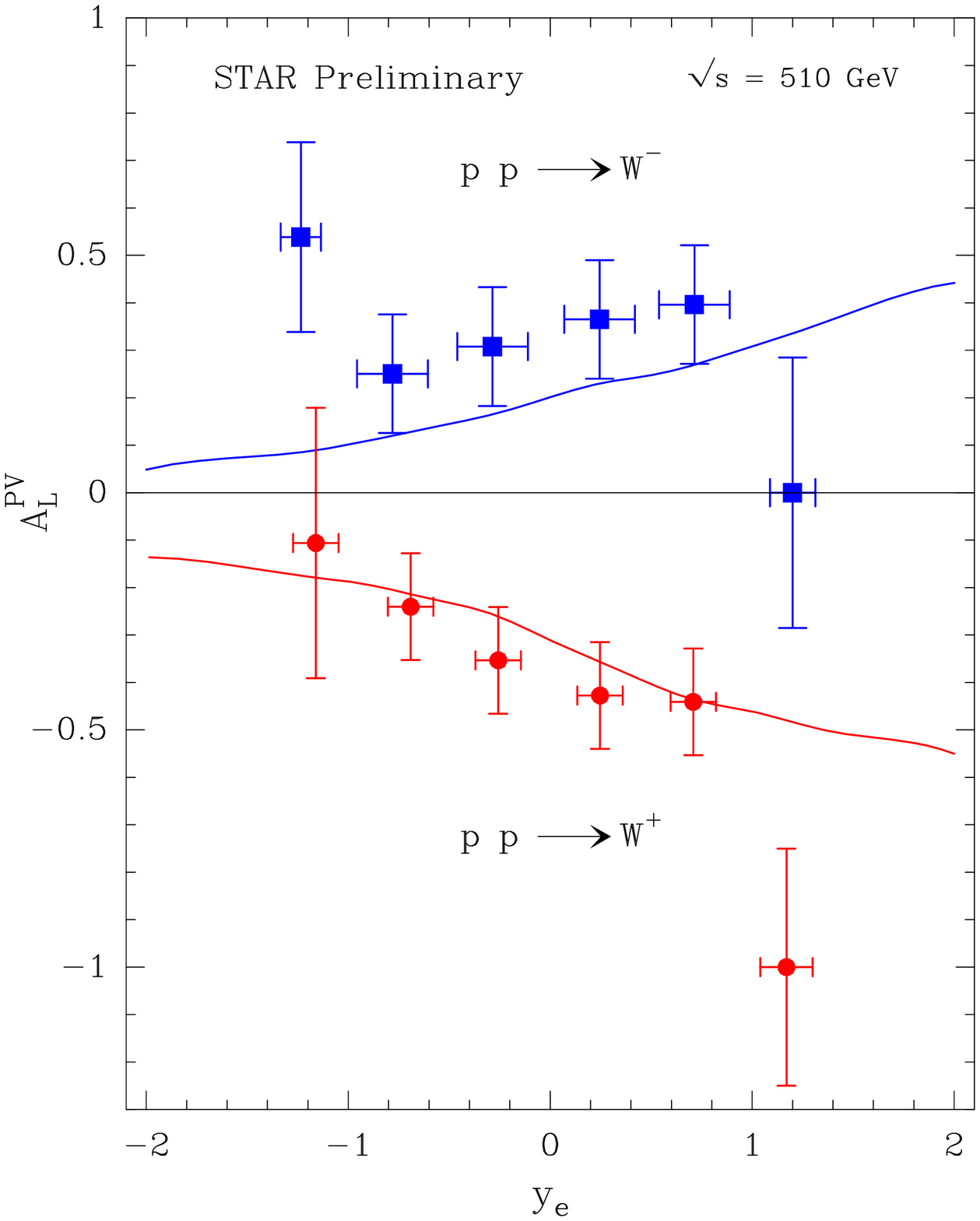}
\end{center}
\vspace*{-8mm}
\caption{{\it Left}: Comparison of the $g_1^{n}(x)$ data at low $Q^2$ from \cite{kramer} with the prediction
of the statistical model. {\it Right}: Predicted parity-violating asymmetries $A_L^{PV}$ for charged-lepton production at BNL-RHIC, through production and decay of $W^{\pm}$ bosons. $y_e$ is the the charged-lepton rapidity and the data points are from Ref. \cite{surrow} (Taken from \cite{bbs13}).
}
\label{fi:fig2}
\end{figure}
 The subject of the strange quark distributions is also very important and challenging, in particular because the HERMES Collaboration has
presented recently a new data set at variance with the previous one. For lack of space we are unable to cover it here.

We now turn to two specific examples of spin-dependent observables to illustrate the predictive power of our approach for helicity quark and antiquark distributions. First, let us consider the neutron structure function $g_1^n(x,Q^2)$ measured in polarized DIS with a neutron target. Although it has been measured extensively by
different collaborations, some accurate data obtained at JLab, in the low $Q^2$ region, have been largely ignored so far \cite{kramer}. In Fig. 2(Left) we compare our predictions with these data, dominated by $\Delta d$ and $\Delta \bar d$ which are negative, and one observes a remarkable agreement. Another example is the helicity asymmetry in the charged-lepton production through production and decay of $W^{\pm}$ bosons. As explained in Ref. \cite{bbs13}, the $W^-$ asymmetry is very sensitive
to the sign and magnitude of $\Delta \bar u$ and the succeful results of the statistical approach are displayed in Fig. 2(Right).

\section{Transverse momentum dependence of the parton distributions}
The parton distributions $p_i(x,k^2_T)$ of momentum $k_T$, must obey the momentum sum rule\\
$\sum_i \int_0^1dx \int x p_i(x,k^2_T) dk^2_T = 1$. In addition it must also obey
the transverse energy sum rule $\sum_i \int_0^1dx \int p_i(x,k^2_T)\frac{k^2_T}{x}dk^2_T =M^2 $.
From the general method of statistical thermodynamics we are led to put $p_i(x,k^2_T)$ in correspondance with the following expression
$\exp({\frac{-x}{\bar{x}}}+{\frac{-k^2_T}{x \mu^2}})$~,
where $\mu^2$ is a parameter interpreted as the transverse temperature.
 So we have now the main elements for the extension to the TMD of the statistical PDF. We obtain in a natural way the Gaussian shape with {\bf no} $x,k_T$ factorization,
because the quantum statistical distributions for quarks and antiquarks read in this case
\begin{equation}
xq^{h}(x,k_T^{2})=\frac{F(x)}{\exp(x-X^{h}_{0q})/\bar{x}+1}
\frac{1}{\exp(k^2_T/x\mu^2-Y^{h}_{0q})+1}~,
\end{equation}
\begin{equation}
x\bar{q}^{h}(x,k_T^{2})=\frac{{\bar
 F}(x)}{\exp(x+X^{-h}_{0q})/{\bar{x}}+1}
\frac{1}{\exp(k^2_T/x\mu^2+Y^{-h}_{0q})+1}~.
\end{equation} 
Here $F(x) = \frac{A x^{b-1}X^{h}_{0q}}{\mbox{ln}(1 + \exp{Y^{h}_{0q}})\mu^2}=\frac{A x^{b-1}}{k\mu^2}$,
where $Y^h_{0q}$ are the thermodynamical potentials chosen such that $\mbox{ln}(1 + \exp{Y^{h}_{0q}})=k X^h_{0q}$,
in order to recover the factors $X^h_{0q}$ and $(X^h_{0q})^{-1}$, introduced earlier.\\
Similarly for $\bar q$ we have $\bar F(x)= \bar A x^{2b-1}/k\mu^2$. The determination of the 4 potentials $Y^h_{0q}$ can be achieved with the choice $k=3.05$. 
Finally $\mu^2$ will be obtained from the transverse energy sum rule and one finds $\mu^2=0.198\mbox{GeV}^2$. Detailed results are shown in Refs.~\cite{bbs6,bbs5}.

\begin{figure}[htp]
\begin{center}
\includegraphics[width=6.5cm]{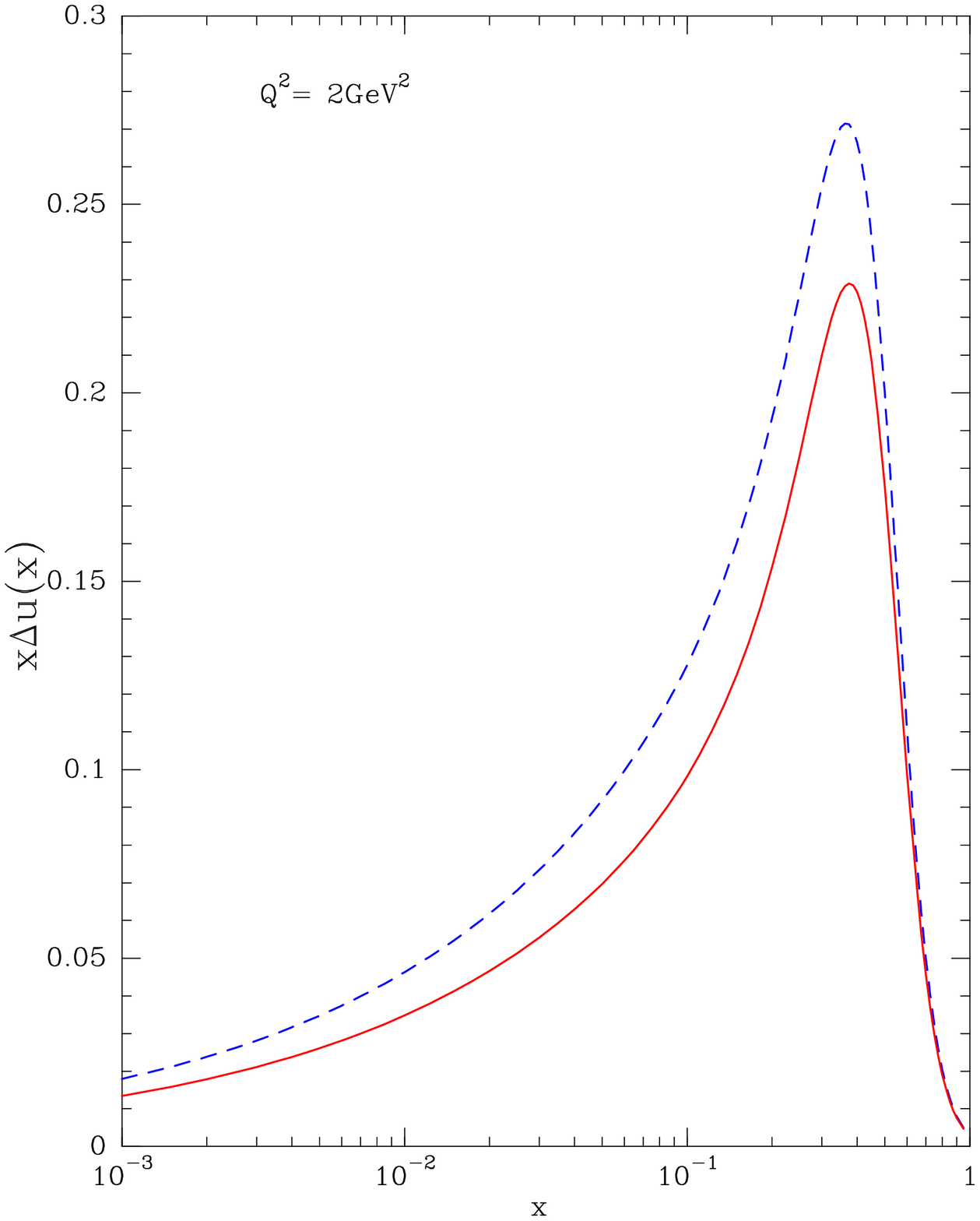}
\vspace*{10mm}
\hspace*{10mm}
\includegraphics[width=6.5cm]{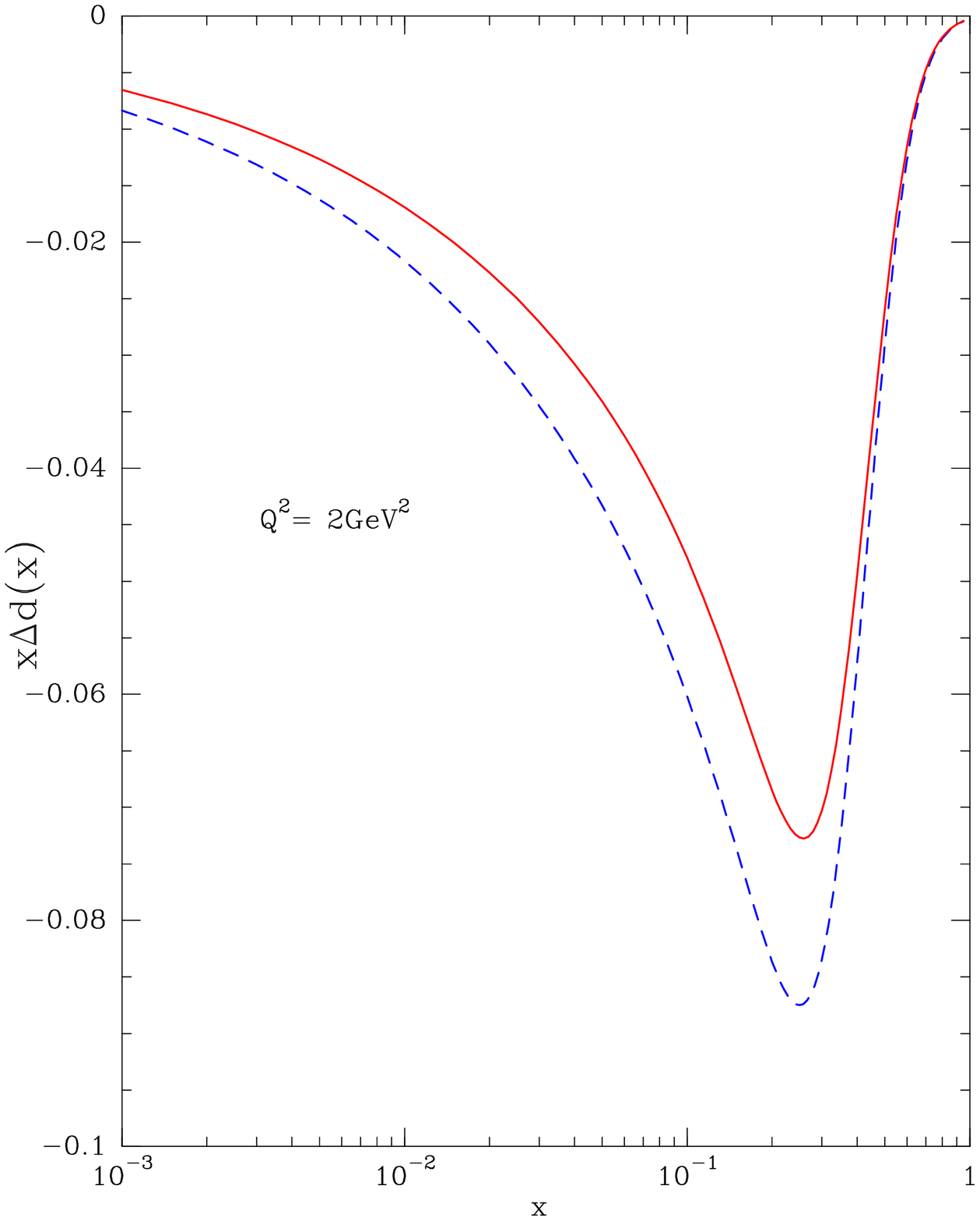}
\end{center}
\vspace*{-8mm}
\caption{The $u$ and $d$ quark helicity distributions versus $x$: $x\Delta q(x)$ ({\it dashed line}) and $x\Delta q^{MW}(x)$ ({\it solid line}). (Taken from Ref.~\cite{bbs5}).}
\label{fi:fig3}
\end{figure}
Before closing we would like to mention an important point.
So far in all our quark or antiquark TMD distributions, the label "`$h$"' stands for the
helicity along the longitudinal momentum and not along the direction of the momentum, as normally
defined for a genuine helicity. The basic effect of a transverse momentum $k_T
\neq 0$ is the Melosh-Wigner rotation, which mixes the
components $q^{\pm}$ in the following way
$q^{+MW}= \cos^2\theta ~q^+ + \sin^2\theta ~q^- ~~~\mbox{and}~~~q^{-MW}=
\cos^2\theta ~q^- + \sin^2\theta ~q^+$, where for massless partons,
$\theta = \arctan{(\frac{k_T}{p_0 +p_z})}$, with $p_0 = \sqrt{k_T^2 +p_z^2}$.
It vanishes when either $k_T =0$ or $p_z$, the quark longitudinal momentum, goes to infinity.
Consequently $q = q^+ + q^-$ remains unchanged since $q^{MW}=q$,
 whereas we have $\Delta q^{MW}= (\mbox{cos}^2\theta - \mbox{sin}^2\theta) \Delta q$.\\ For illustration we display in Fig. 3, $x\Delta q(x)$ and $x\Delta q^{MW}(x)$ for
$Q^2 = 2 \mbox{GeV}^2$, which shows the effect of the Melosh-Wigner rotation, mainly in the low $x$ region.\\

A new set of PDF is constructed in the framework of a statistical approach of the nucleon.
All unpolarized and polarized distributions depend upon a small number of
free parameters, with some physical meaning.
New tests against experimental (unpolarized and polarized)
data on DIS, semi-inclusive DIS and hadronic processes are very satisfactory.
It has a good predictive power but some special features remain to be verified, specially in the high $x$ region.
The extension to TMD has been achieved and must be checked more accurately together with Melosh-Wigner effects in the low $x$ region, for small $Q^2$.\\ 

{\bf Acknowledgments}\\
JS is grateful to the organizers of DSPIN-13 for their warm hospitality at JINR and for their invitation to present this talk. Special thanks go to Prof. A.V. Efremov for providing a full financial support and for making, once more, this meeting so successful.

\end{document}